\documentclass[aps,twocolumn,amsmath,amssymb,prb,showpacs,superscriptaddress]{revtex4-1}
\usepackage{amsfonts, hyperref, times}
\usepackage{graphicx}
\usepackage{color}
\usepackage[english]{babel}

\providecommand{\vect}[1]{{\boldsymbol{#1}}}

\begin{document}
  
\title{Facilitating domain wall injection in magnetic nanowires by electrical means}

\author{Davi~R. Rodrigues}
\affiliation{Institute of Physics, Johannes Gutenberg University of Mainz, 55128 Mainz, Germany}
\author{Nils Sommer}
\affiliation{Forschungszentrum J\"ulich, Peter Gr\"unberg Institut, 52425 J\"ulich}
\author{Karin Everschor-Sitte}
\affiliation{Institute of Physics, Johannes Gutenberg University of Mainz, 55128 Mainz, Germany}

\date{\today}

\begin{abstract}

We investigate how to facilitate the injection of domain walls in chiral ferromagnetic nanowires by electrical means. 
We calculate the critical current density above which domain walls are injected into the nanowire 
depending on the material parameters and the source of interaction including spin-transfer torques as well as spin-orbit torques.
We demonstrate that the Dzyaloshinskii-Moriya interaction can significantly reduce the required critical current to inject the types of domain walls favored by the Dzyaloshinskii-Moriya interaction. 
We find that in chiral magnets it is only possible to shed a single domain wall by means of spin-orbit torques, as they modify the ground state orientation of the system. In contrast, for spin-transfer-torque induced shedding of domain walls, we show that there exist two different critical current densities for the two different domain wall chiralities, respectively. 
 Additionally, for 
the consecutive creation of domain walls by means of spin-transfer torques, we find that the interaction between the domain walls cannot be neglected and even may lead to the pairwise annihilation of consecutive domain walls with opposite chiralities. 
 \end{abstract}


\maketitle

%%%%%%%%%%%%%%%%%%%%%%%%%%%%%%%%%%%%%%%%%%%%%%%%%%%%%%%%%%%%%%%%%%%%%%%%%%%%%%%%%%
%%%%%%%%%%%% Section I
%%%%%%%%%%%%%%%%%%%%%%%%%%%%%%%%%%%%%%%%%%%%%%%%%%%%%%%%%%%%%%%%%%%%%%%%%%%%%%%%%%
\section{Introduction}
(Meta)stable magnetic configurations in ferromagnetic materials have attracted a lot of attention due to their promising applications for spintronic devices.~\cite{Kosevich1990, Stamps2014, Sander2017, Everschor-Sitte2018, Back2020} 
In particular, due to their particle like behavior and stability,~\cite{Schryer1974,Yamaguchi2004, Klaui2008, Rodrigues2017} 
magnetic domain walls are considered as key elements for (potential) spintronics based devices such as 
magnetic sensors,~\cite{Weiss2013} 
the race track memory,\cite{Parkin2008, Hayashi2008a} 
magnetic logic,\cite{Allwood2005} 
domain-wall-based magnonic nanocircuitry,\cite{Wagner2016}
or the implementations of artificial neurons~\cite{Sengupta2016} and synapses~\cite{Sharad2012}.
 Therefore, controllable low power electrical means are needed to create magnetic domain walls.~\cite{Ravelosona2006,Hayashi2008a,Chanthbouala2011,Sitte2016,Dao2019}

\begin{figure}
\includegraphics[width=1.0 \columnwidth]{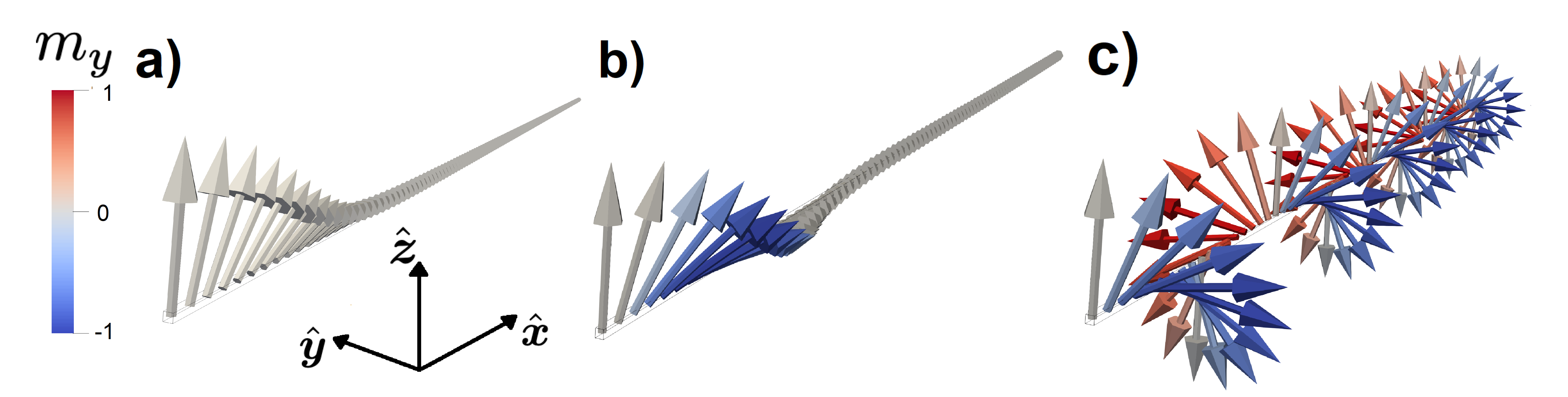}
\caption{Ground states of the magnetic semi-infinite nanowire with fixed magnetization direction at the boundary for different DMI strengths. In a) there is no DMI, in b) and c)  the DMI strength is below and above the critical value  $D_c^2 = 2 J \lambda$, respectively. The chiral interaction induces an additional twist of the magnetization along the wire. Above the critical value, a helical state forms the ground state.~\cite{Bode2007,Uchida2006,Rohart2013} The colors represents the $y$ component of the magnetization.}
\label{fig:groundstates_without_with_and_with_too_high_DMI}
\end{figure}

In ferromagnetic materials domain walls can be injected electrically at inhomogeneities which for example occur naturally at the edges of a magnetic nanowire.~\cite{Sitte2016}  
The generic principle behind this creation mechanism is that a spin-polarized current exerts spin-torques on the magnetic system, thereby, in particular, inducing a twisted state in the inhomogeneity region, see Fig.~\ref{fig:groundstates_without_with_and_with_too_high_DMI}b).
Above a certain threshold current density $j_c$, the current induced spin-torques will be so strong that the 
 twisted generated spin structure will tear off and travel dynamically along the wire.~\cite{Ravelosona2006,Hayashi2008a,Chanthbouala2011,Sitte2016,Dao2019}
Note that this fundamental principle is independent of detailed microscopic mechanisms such as the origin of the inhomogeneity, the source of the spin-torques, etc. 
For example, the periodic domain wall injection by means of spin-transfer torques (STTs)~\cite{Slonczewski1996,Berger1996,Stiles2002} has been predicted in a simple model considering exchange and anisotropy interaction only. In this model, the critical current density as well as the magnetization profiles have been calculated, and the production period was shown to behave as $T \sim (j-j_{c})^{-1/2}$ with a universal exponent being independent of micromagnetic details.~\cite{Sitte2016} 
Furthermore, recently the injection of domain walls via spin-orbit torques (SOTs)~\cite{Miron2011,Ryu2013} was experimentally observed in nanomagnets subject to chiral interactions.\cite{Dao2019}  A similar design has also been considered to inject domain walls in ferromagnetic insulators using magnons.~\cite{Hartmann2020}
 
Magnets in which inversion symmetry is broken, typically allow for chiral interactions such as the Dzyaloshinskii-Moriya interaction.~\cite{Dzyaloshinsky1958,Moriya1960a}  Besides stabilizing a rich set of chiral spin structures,\cite{Bogdanov1989, Uchida2006, Bode2007, Mohseni2013, Ryu2013, Everschor-Sitte2018} chiral interactions allow for the required synchronous motion of multiple domain walls by magnetic fields.\cite{Gambardella2011, Ryu2013, Emori2013,Kim2014b}

In this work, we provide a detailed theory for the injection threshold current of domain walls in chiral magnets including the effects of STTs and SOTs.
We show that twisting terms, such as chiral interactions, simplify the injection of domain walls in the sense that a they reduce the critical current density needed for domain wall creation. We find that for SOTs, it is only possible to inject a single domain wall. For injecting another domain wall, switching the main magnetization direction along the wire back to its original state is required first.
For the STT-induced creation, we find a domain-wall-chirality dependent critical current density. This causes a difference in the period for the creation of two consecutive domain walls with opposite chirality. We further study the dynamics of the shedded domain walls and derive a simple model in terms of their collective coordinates for their mutual interaction.  

This paper is organised as follows: In Sec.~\ref{sec:Model} we present the model for a semi-infinite chiral ferromagnetic nanowire subject to current-induced spin-torques.
In Sec.~\ref{sec:CreationDW} we derive the threshold currents above which domain walls are injected into the nanowire, taking into account the DMI, STTs and SOTs. In Sec.~\ref{sec:DWInteraction} we study the domain wall dynamics of shedded domain walls. We derive the asymptotic interaction of domain walls at long distances, which lead to an oscillatory yo-yo effect like distance dependence between the moving domain walls. 
Finally, we summarize our results in Sec.~\ref{sec:Conclusion}

%%%%%%%%%%%%%%%%%%%%%%%%%%%%%%%%%%%%%%%%%%%%%%%%%%%%%%%%%%%%%%%%%%%%%%%%%%%%%%%%%%
%%%%%%%%%%%% Section II 
%%%%%%%%%%%%%%%%%%%%%%%%%%%%%%%%%%%%%%%%%%%%%%%%%%%%%%%%%%%%%%%%%%%%%%%%%%%%%%%%%%
\section{Micromagnetic model}
\label{sec:Model}
For the injection of domain walls in nanowires we consider a semi-infinite ferromagnetic nanowire with an easy axis along the wire, see Fig.~\ref{fig:groundstates_without_with_and_with_too_high_DMI}. The magnetization at one end of the wire is pinned perpendicular to the easy axis, which we will denote as $\hat{\vect z}$ axis in the following. The pinning may be due to a local inhomogeneity or an external local interaction, such as a strong magnetic field. The energy for this system is given by~\cite{Tretiakov2010}
\begin{equation}
E\left[\vect{m}\right] = \int_0^\infty dx \Bigl[\frac{J}{2} \left(\partial \vect{m}\right)^2 + \lambda (1-m_x^2) + D \vect{m} \cdot \left(\hat{\vect{x}}\times \partial\vect{m}\right)\Bigr],
\label{eq:energy}
\end{equation}
where $\vect{m} = \vect{M}/M_{s}$ is the unitary magnetization and $M_{s}$ is the saturation magnetization, $J$, $\lambda$ and $D$ are the exchange, anisotropy and Bloch DMI strengths respectively, and $\partial \equiv \partial /\partial x$.
The energy associated to the presence of a single chiral domain wall with domain wall width $\Delta = J/\sqrt{2J\lambda- D^2}$ is given by $E_{\textrm{chiralDW} }= 2J/\Delta$.~\cite{Tretiakov2010} For a DMI strength larger than a critical value, $D > D_{c} \equiv \sqrt{2J\lambda}$, the ground-state changes from a ferromagnetic to a helical state,~\cite{Bode2007,Uchida2006,Rohart2013} as the DW energy becomes smaller than the energy of the ferromagnet which is zero in this model, see Fig.~\ref{fig:groundstates_without_with_and_with_too_high_DMI}.
As the focus of this work is to study domain wall creation in a ferromagnetic background, we consider $D < D_c$ in the following.

The dynamics of the magnetization in the presence of a spin polarized current is given by the Landau-Lifshitz-Gilbert-Slonczewski (LLGS) equation ~\cite{Slonczewski1996}
\begin{equation}
\vect{\dot{m}} = - \gamma \vect{m} \times \vect{H}_\mathrm{eff} + \alpha \vect{m} \times \vect{\dot{m}} - \vect{\tau}_\mathrm{STT} - \vect{\tau}_\mathrm{SOT}, \label{eq:LLG}
\end{equation}
where $\gamma$ is the gyromagnetic ratio, $\alpha$ the Gilbert damping parameter, and 
\begin{align}
\vect{H}_\mathrm{eff} =& -\frac{1}{M_{s}}\left(\frac{\delta E \left[\vect{m}\right]}{\delta \vect{m}}\right)\notag\\
 =&\frac{1}{M_{s}}\left( J \partial^2\vect{m} - 2 D \hat{\vect{x}}\times\partial\vect{m} + 2 \lambda\, m_{x} \hat{\vect{x}} \right)
\end{align}
is the effective magnetic field.
The STT and SOT terms 
are given by~\cite{Slonczewski2002,Thiaville2005,Zhang2004,Garate2010,Hayashi2014}
\begin{subequations}
\begin{align}
\vect{\tau}_\mathrm{STT} &= v\, \partial \vect{m} - \beta v \, \vect{m} \times \partial \vect{m}, \label{eq:STT_torque}\\
\vect{\tau}_\mathrm{SOT} &= \tau_{\mathrm{DL}} \left(\xi \vect{m} \times \vect{\sigma} + \vect{m} \times \left(\vect{m} \times \vect{\sigma}\right)\right)
\label{eq:SOT_torque}.
\end{align}
\end{subequations}
Both torque terms constitute of a field-like term, which conserves energy, and a non-conservative damping-like term. By convention, one denotes by $\beta$ the ratio between the damping-like and field-like terms for STTs and similarly by $1/\xi$ for SOTs.~\cite{Zhang2004,Garate2010,Hayashi2014}
These torques are induced by spin-currents which have different origins.

For STTs~\cite{Baibich1988,Berger1996,Slonczewski1996,Stiles2002,Zhang2004}   an electric current passes through a ferromagnetic material and thus becomes spin-polarised. 
The resulting spin velocity $v$ is proportional to the electric current density $j^{\textrm{STT}}$ and given by ~\cite{Zhang2004}
\begin{equation}\label{eq:Zhang}
v = \frac{P \mu_B}{e M_s\left(1+\beta^2\right)} j^{\textrm{STT}}\: ,
\end{equation}
where $P$ is the current polarization, $\mu_B$ is the Bohr magneton, and $e$ is the electron charge.
We consider $v$ to be positive, to allow for shedding of domain walls along the wire.

SOTs are induced by the spin Hall\cite{Dyakonov1971b,Zhang2000} or the Rashba-Edelstein\cite{Bychkov1984,Edelstein1990} effect. These effects occur naturally at interfaces between ferromagnets and heavy metals or topological insulators.~\cite{Geranton2015,Mahfouzi2012a} Here the spin polarized current is generated perpendicular to the electrical current and the normal direction $\hat{\vect{n}}$ of the interface between the materials, $\vect{\sigma} = \hat{\vect{n}}\times\vect{j}^{\textrm{SOT}}$.  The proportionality constant $\tau_{\mathrm{DL}}$ depends on the details of the materials as\cite{Hayashi2014}
\begin{align}
\tau_{\mathrm{DL}} = \frac{\gamma \hbar \,\theta_\mathrm{Hall}}{2 M_s e\, l}\label{eq:tau_DL}
\end{align}
where $\hbar$ is the reduced Planck constant, $\theta_\mathrm{Hall}$ is the spin Hall ratio, and $l$ the thickness of the ferromagnetic layer.

%%%%%%%%%%%%%%%%%%%%%%%%%%%%%%%%%%%%%%%%%%%%%%%%%%%%%%%%%%%%%%%%%%%%%%%%%%%%%%%%%%
%%%%%%%%%%%% Section III 
%%%%%%%%%%%%%%%%%%%%%%%%%%%%%%%%%%%%%%%%%%%%%%%%%%%%%%%%%%%%%%%%%%%%%%%%%%%%%%%%%%
\section{Creation of domain walls by spin currents in chiral magnetic wires}
In this section we present our analytical and numerical results for the domain wall creation by spin currents in chiral magnetic nanowires subject to STTs or SOTs, and provide the derivation of the threshold current densities above which domain wall injection takes place. 

For the numerical parts, we considered a wire with $1024$ lattice sites separated at $3$ nm each. 
To lock the direction of the magnetization at the boundary, we applied a strong magnetic field, $B_{\textrm{ext}} =10^7$ A/m, to the first $100$ lattice sites, along the $z$-direction, i.e.\ perpendicular to the direction of the wire. The micromagnetic simulations were performed using an internal version of MicroMagnum~\cite{MicroMagnum}, with magnetization saturation $M_{s} = 6.0\times10^5$ A/m, exchange constant $J = 2.6\times10^{-11}$ J/m, uniaxial anisotropy strength $\lambda = 1.0\times10^4$ J/m$^3$, if not explicitly stated otherwise. For the STTs we used a spin current polarization of $P = 0.56$ and for the SOTs we used $\theta_\mathrm{Hall} = 1$. 
To numerically obtain the threshold current densities, we employed the method of nesting intervals, i.e.\ we screened the current strengths for the injection of domain walls with fixed material parameters, thereby confining the interval in which the critical current density is located. The intervals in our numerical simulations corresponded to $\lesssim 0.1\%$ of the expected analytically calculated value.

\subsection{Spin-transfer torques}
\label{subsec:STT}

\label{sec:CreationDW}
\begin{figure}
\includegraphics[width=1.0 \columnwidth]{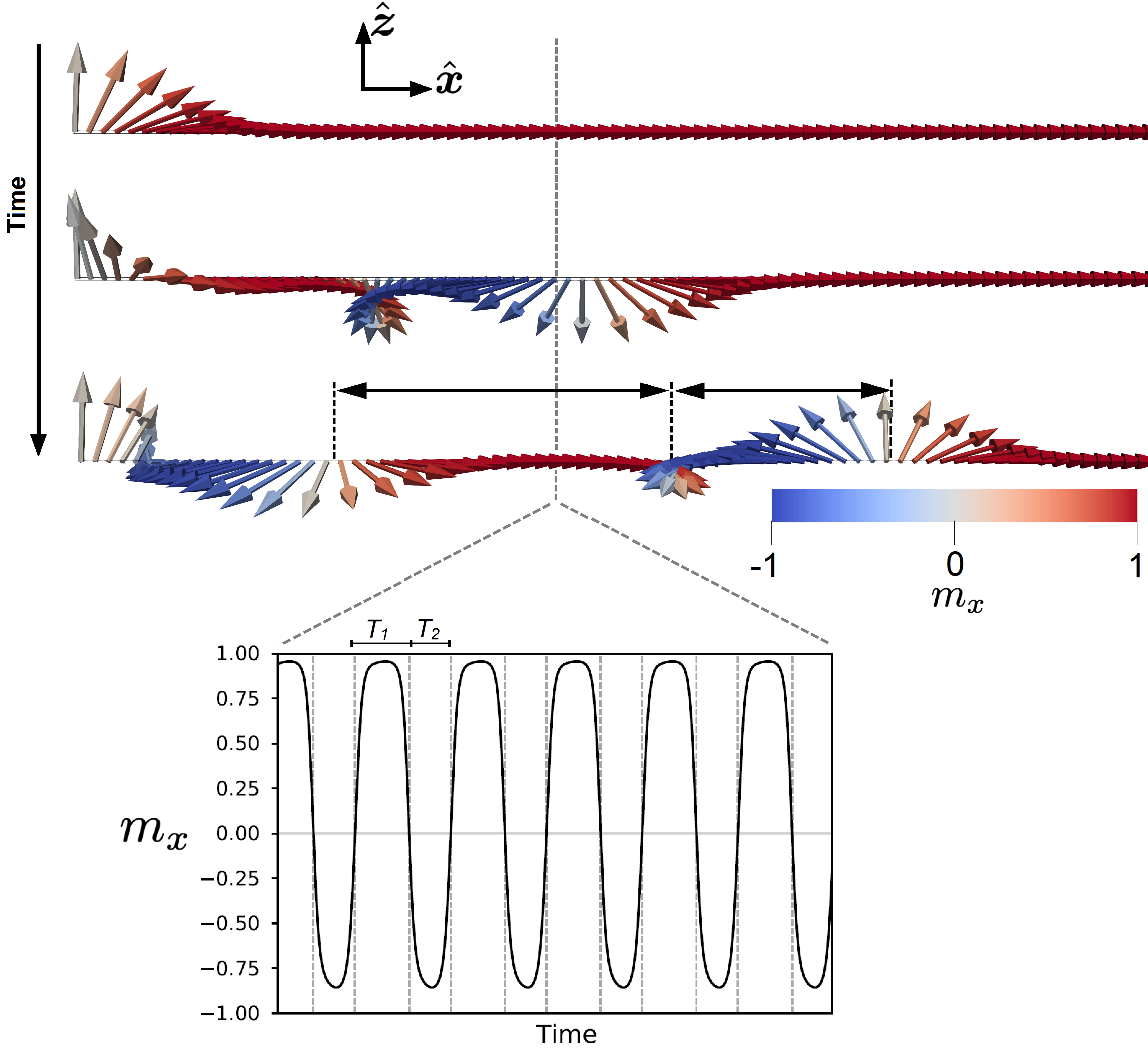}
\caption{Depiction of periodic domain wall creation by STTs in a semi-infinite chiral nanowire. The color code represents the $x$ component of the magnetization. The shedding periods are different for tail-to-tail ($T_1$) or head-to-head ($T_2$) domain walls, due to the DMI.
The distances between two consecutive domain walls are initially proportional to the period of creation, however, change in time due to the different velocities of consecutive chiral domain walls, see Sec.~\ref{sec:DWInteraction}. The graph in the bottom shows the time evolution of the $m_x$ component of the magnetization over time for a single point of the nano-wire away from the origin.
}
\label{fig:SheddingDW}
\end{figure}

STTs allow for the periodic creation of domain walls, which in the presence of DMI, leads to two different shedding periodicities, see Fig.~\ref{fig:SheddingDW}.
Within this part we derive the STTs induced critical current density $j_c^{\mathrm{STT}}$ by similar means as developed in Ref.~\citenum{Sitte2016}. 
Projecting Eq.~\eqref{eq:LLG}, i) into the direction along the wire, $\hat{\vect{x}}$, 
 and ii) onto $\vect{m}\times \partial \vect{m}$ yields the following two conditions for the existence of a stable static solution,
\begin{subequations}\label{eq:halfdomainmomentaSTT}
\begin{align}
&\partial\Bigl(J\hat{\vect{x}}\cdot\left(\vect{m}\times\partial\vect{m}\right) + D \bigl(m_{x} + \frac{v M_{s}}{2D\gamma}\bigr)^2\Bigr) =\notag\\
&\qquad \qquad \qquad \qquad \quad \frac{\beta v M_{s}}{\gamma} \hat{\vect{x}}\cdot\left(\vect{m}\times\partial\vect{m}\right),\\
&\partial\left(\frac{J}{2} (\partial\vect{m})^2 + \lambda m_{x}^2\right)  = \frac{\beta v M_{s}}{\gamma} (\partial\vect{m})^2.
\end{align}
\end{subequations}
The above expressions on the left hand side, correspond to the angular and linear momenta of the magnetic configuration.
As expected, in the presence of a non-conservative torque (i.e.\ $\beta \neq0$), neither the linear momentum nor the angular momentum are conserved along the wire direction.

\begin{figure}
\includegraphics[width = 0.9 \columnwidth]{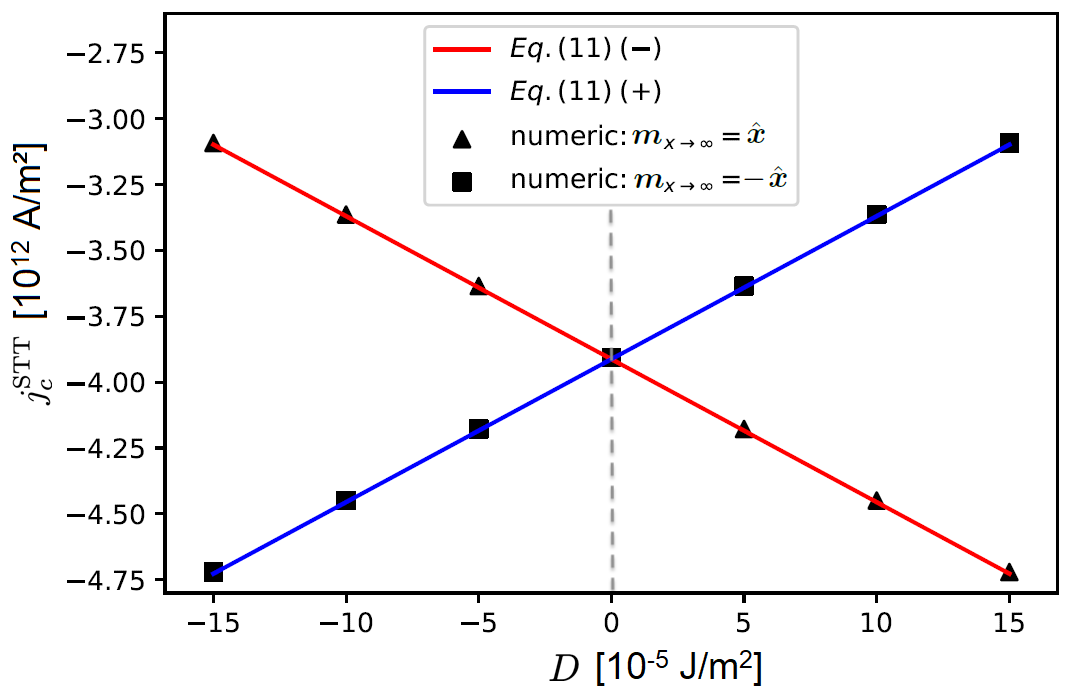}
\caption{Critical current density for domain wall creation by STTs for $\beta =0$ as a function of DMI strength.
The DMI energetically favours one type of domain wall, corresponding to different ground states at infinity. Thus, while it facilitates the creation by lower current densities, it enhances it for the other domain wall type. The grey dashed line corresponds to $D = 0$, when the energy of both types of domain walls are the same.}
\label{fig:STTwithDMI}
\end{figure}

\subsubsection{Limiting case: Absence of non-adiabatic torques}
For simplicity, we first consider $\beta = 0$, where the two momenta are conserved, and the system is fully analytically solvable. Comparing the expressions of the respective conserved momenta at both ends of the wire, i.e.\
at $x=0$ where $\vect{m} = \hat{\vect{z}}$,
and 
at $x \rightarrow \infty$ where the $\vect{m} = \pm\hat{\vect{x}}$ and $\partial \vect m =0$, 
we obtain 
\begin{subequations}\label{eq:solutions_of_STT}
\begin{align}\label{eq:solutions_of_STTa}
-\partial m_y|_{x=0} &= \frac{D}{J} \pm v \frac{ M_s}{\gamma  J}\\
\label{eq:solutions_of_STTb}
\left(\partial \vect{m}\right)^2|_{x=0} &= \frac{2\lambda}{J}.
\end{align}
\end{subequations} 
Note that the sign in Eq.~\eqref{eq:solutions_of_STTa} depends on the direction of $\vect{m}$ at infinity.
Furthermore, exploiting the general relation $\vect{m}\perp \partial\vect{m}$ at $x=0$ yields  $\left(\partial \vect{m}\right)^2|_{x=0} = \left(\partial m_x\right)^2|_{x=0} +\left(\partial m_y\right)^2|_{x=0}$. 
By combining $0< \left(\partial m_x\right)^2 = 2\lambda/J - \left(\partial m_y\right)^2$ with Eq.~\eqref{eq:solutions_of_STTa}, we obtain as a condition for having a static solution
\begin{equation}
\Bigl(\frac{M_s}{\gamma}v \pm D\Bigr)^2 < 2 J \lambda.
\end{equation}
 Note that for $v=0$ this equation is automatically fulfilled in the ferromagnetic state where $D < D_c = \sqrt{2 J\lambda}$. 
The equation above demonstrates that a static solution exists only for currents below the critical spin velocity $v_c$ given by 
\begin{equation}
v_c = \frac{\gamma}{M_s} \left(\sqrt{2 J \lambda} \mp D\right),
\end{equation}
corresponding to the critical current density
\begin{equation}
j_c^{\textrm{STT}} = \frac{e M_s}{P \mu_B}v_c = \frac{\gamma e }{P \mu_B} \left(\sqrt{2 J \lambda} \mp D\right).
\label{eq:Solution_STT}
\end{equation}
The different signs correspond to the different types of domain walls, head-to-head and tail-to-tail. Notice that each of these types will have a different sense of rotation, i.e.\ helicity, given by the DMI. Since they have different energies due to the chiral interaction, for a given current strength, they require different amounts of time to be injected into the wire. 
We have confirmed our analytic calculations by means of micromagnetic simulations. In particular, we find that the critical current depends linearly on the DMI strength, $D$, up to the point when the system decays to the helicoidal state, as shown in Fig.~\ref{fig:STTwithDMI}.

\subsubsection{Including non-adiabatic torques}
\begin{figure}
\includegraphics[width = 1.0 \columnwidth]{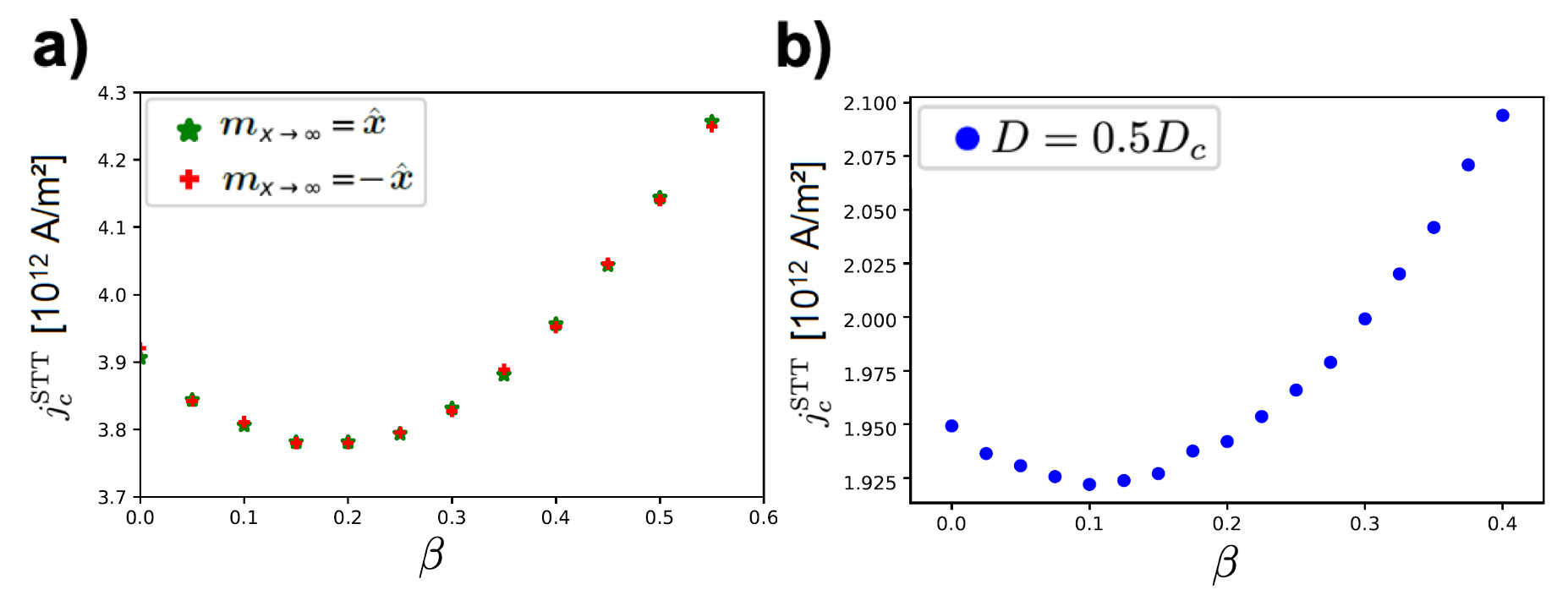}
\caption{Critical current density as a function of the non-adiabatic spin-torque strength $\beta$ for a) both possible ground states at infinity for $D=0$ and b) for $\vect{m} = \hat{\vect{x}}$ at infinity and $D=0$.
The critical current density has a global minimum value for $\beta > 0$.
In a) we observe that the dependence on $\beta$ is the same for both ground states at infinity. In b) we show that $j_{c}^{\mathrm{STT}}$ is overall smaller than for $D=0$. Also, for $|D| >0$, the position of the minimum shifts to smaller $\beta$.} 
\label{fig:betaSTT}
\end{figure}

To obtain the critical current for the non-conservative case, i.e.\ $\beta \neq 0$, we integrate Eqs.~\eqref{eq:halfdomainmomentaSTT} over the nanowire, see App.~\ref{app:jcrit}. The critical current depends on the material parameters as well on the exact magnetization configuration $\vect{m}$ at the critical current, for the calculation of the profiles see App.~\ref{app:dwprofile}. Here, we illustrate the result for the simpler case, $D=0$, where we obtain 
\begin{multline}
\label{eq:jcSSTbeta}
j_{c}^{\textrm{STT}}(\beta) = \frac{e M_{s} (1+\beta^2)}{P\mu_{B}} \times\\ 
\frac{\beta E_{\textrm{exch}} + \sqrt{\left(\beta E_{\textrm{exch}}\right)^2 + 2\lambda J\left(\pm1 + \beta
E_{\textrm{hel}}
\right)^2}}{\left(1 + \beta |E_{\textrm{hel}}|
\right)^2}
\end{multline}
where $E_{\textrm{exch}} = (J/2)\int_{0}^{\infty}dx\,(\partial\vect{m})^2$ is the exchange energy, and $E_{\textrm{hel}}= \int_{0}^{\infty} dx \, \hat{\vect{x}}\cdot(\vect{m}\times\partial\vect{m})$ the helicity energy of the domain wall texture $\vect m$ which solves Eq.~\eqref{eq:halfdomainmomentaSTT}.
Note that the helicity energy breaks inversion symmetry and thus prefers a rotational sense, as is done by the STT terms which deflect the magnetic state out of the $\vect{x}-\vect{z}$ plane. Furthermore, it has the same structure as the DMI term, $E_{\textrm{hel}} = - E_{\textrm{DMI}}/D$, from which one can gain energy by twisting the magnetic configuration.
By noticing that the integrants of $E_{\textrm{exch}}$ and $E_{\textrm{hel}}$ are always positive for the given boundary conditions, we can obtain the general behavior for the critical current dependence as a function of $\beta$. 
The current initially decreases for small $\beta$ until it reaches its global minimum value. For higher $\beta$ the critical current increases monotonically, see Fig.~\ref{fig:betaSTT}. 

For chiral systems, i.e.\ $D \neq 0$, we find that the critical current density is reduced, as expected, since the system is pre-twisted, see Fig~\eqref{fig:betaSTT}. 
As a function of $\beta$ we find the same trend, first it reduces until it reaches its global minimum and then increases again.
We further observe that the position of the minimum is shifted to smaller $\beta$, see App.~\ref{app:STTdetails}.

\subsection{Spin-orbit torques}
\label{subsec:SOT}
\begin{figure}
\includegraphics[width=0.5 \textwidth]{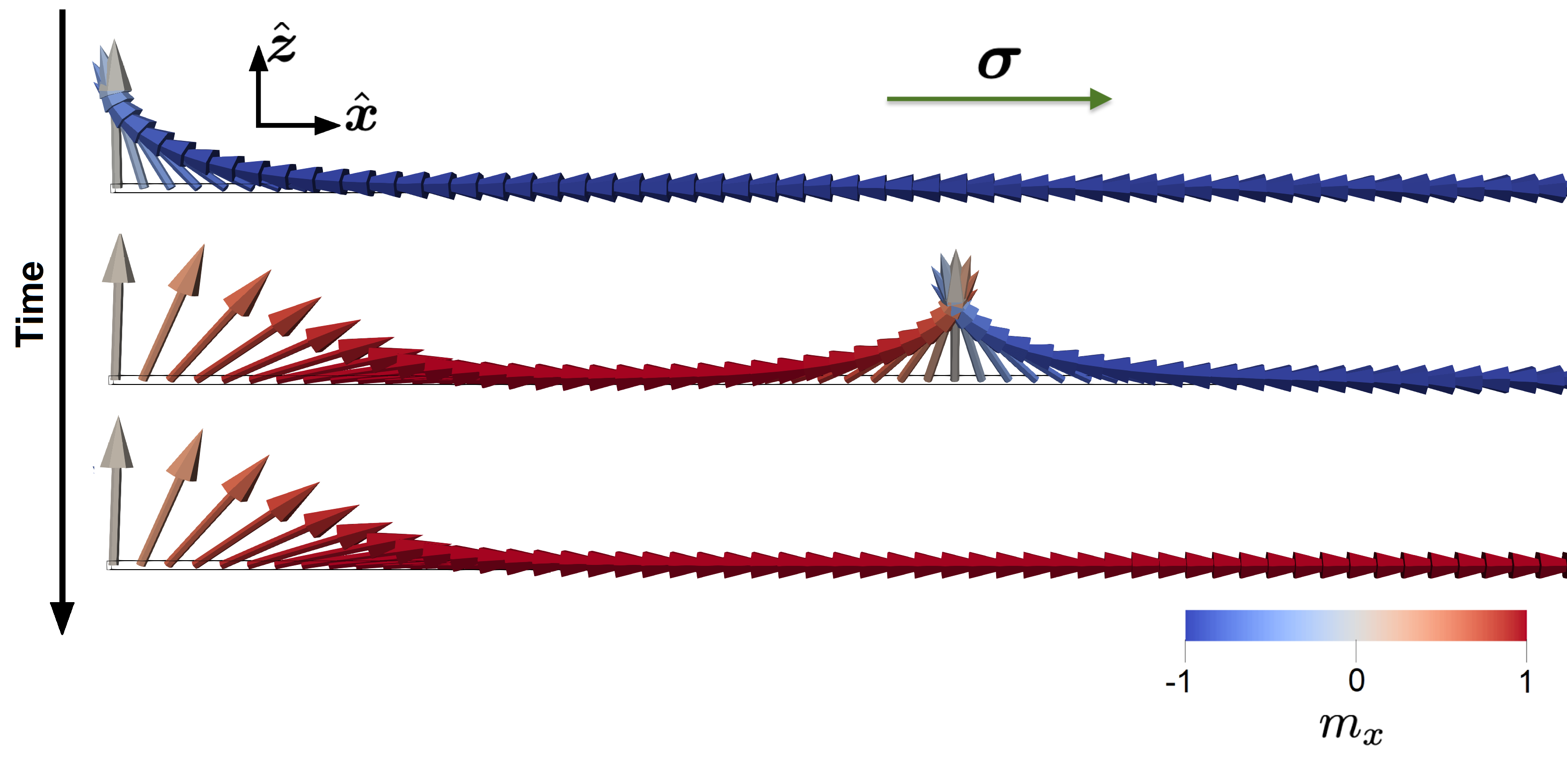}
\caption{Sketch of the domain wall production by SOTs in a semi-infinite chiral nanowire. The color code represents the $x$ component of the magnetization. Initially, at infinity we have the state at $\vect{m} = -\hat{\vect{x}}$, corresponding to the metastable state for a spin current $\vect{\sigma} = j^{\textrm{SOT}}  \hat{\vect{x}}$. After some time, the domain wall is created and travels such that in the end one obtains $\vect{m} = \hat{\vect{x}}$ corresponding to the ground state for the same spin current.
}
\label{fig:SheddingDWSOT}
\end{figure}

SOTs act directly on the magnetization $\vect{m}$, and not on the gradient in contrast to STTs. As a consequence, the presence of SOTs lifts the degeneracy of the ground state of a nanowire with an easy axis, fixing it along the direction of the applied spin current, while the ferromagnetic state in the opposite direction becomes a metastable state.~\cite{Tatara2004,Miron2011,Ryu2013,Dao2019} Thus,
 given a spin current along the easy-axis, it is possible to switch the magnetization from the metastable state, antiparallel to the spin current, to the ground state, parallel to the spin current, and produce a single domain wall, see Fig.~\ref{fig:SheddingDWSOT}.
  To obtain consecutive domain wall creation, it is necessary to either switch the boundary conditions or the direction of the current.~\cite{Dao2019}

In this section we derive the minimal current necessary to inject a domain wall given the setup shown in Fig.~\ref{fig:SheddingDWSOT}.
We consider the spin current along the easy-axis, i.e.\ $\vect{\sigma} = j^{\textrm{SOT}} \vect{\hat{x}}$, and  
an initial metastable ferromagnetic configuration at infinity, $\vect{m} = -\hat{\vect{x}}$.
By the same methodology as in Sec.~\ref{subsec:STT}, we obtain
\begin{subequations}\label{eq:halfdomainmomentaSOT}
\begin{align}
&\partial \left(J(\hat{\vect{x}}\cdot (\vect{m}\times\partial\vect{m})) - D(1- m_{x}^2)\right) =
 j^{\textrm{SOT}} \tilde{\tau} \left(1 - m_{x}^2 \right),\\
&\partial\Bigl(\frac{J}{2}(\partial\vect{m})^2  + \lambda m_{x}^2 + \xi j^{\textrm{SOT}}  \tilde{\tau} m_{x} \Bigr) 
= j^{\textrm{SOT}} \tilde{\tau} \hat{\vect{x}}\cdot(\vect{m}\times\partial\vect{m}),
\end{align}
\end{subequations}
where we have introduced $\tilde{\tau}= \tau_{\mathrm{DL}} M_{s}/\gamma$ to shorten the notation.

\begin{figure}
\includegraphics[width = 1.0\columnwidth]{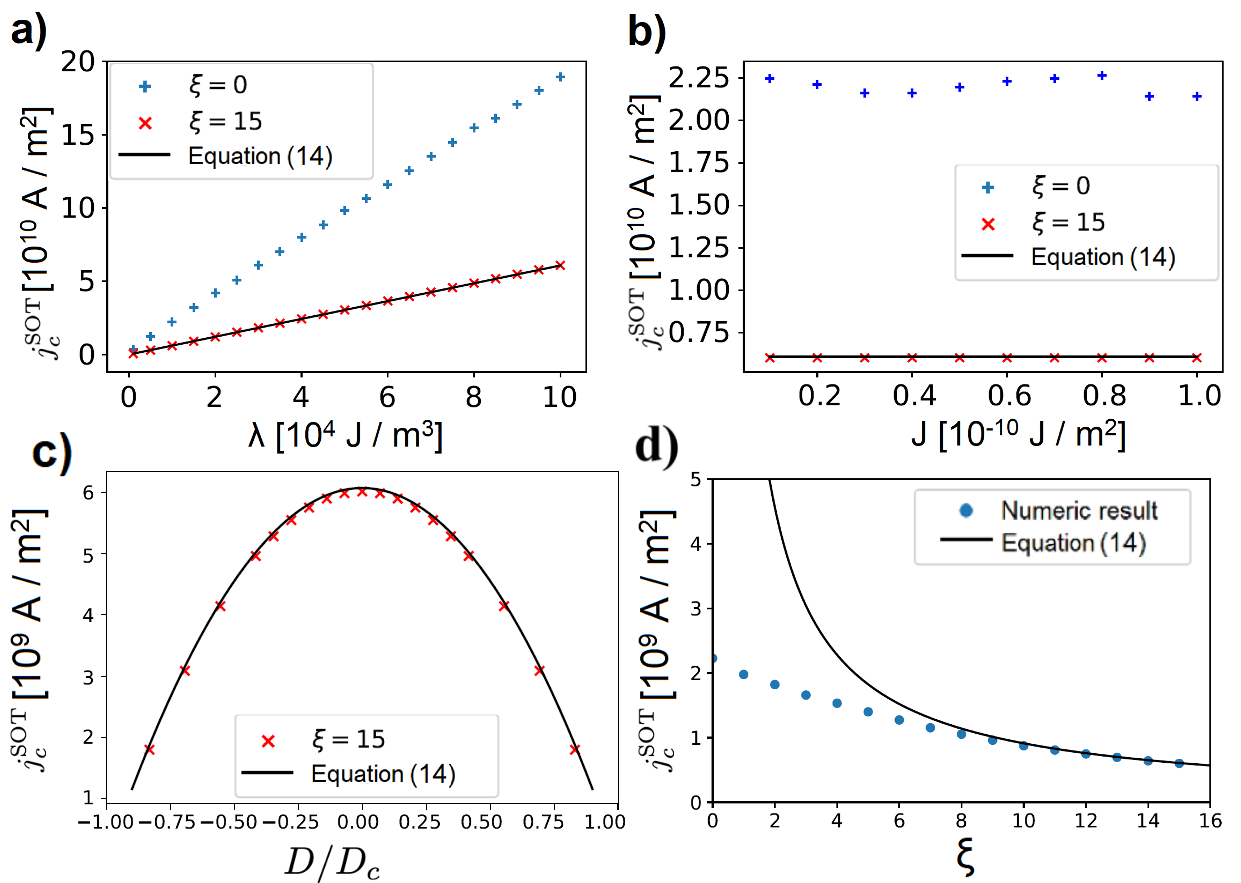}
\caption{Dependence of the critical current density $j_c^{\textrm{SOT}}$ on 
a) the anisotropy strength $\lambda$ for $D=0$;
b) the exchange interaction strength $J$ for $D=0$;
c) the DMI strength for 
   $\xi = 15$; and 
d) the ratio of field vs.\ damping like torque $\xi$ for $D=0$.
The other parameters are chosen to be $l = 3\, nm$ and $\theta_\mathrm{Hall} = 1$.
The analytic solution, Eq.~\eqref{eq:result_SOT}, which is valid in the limit $\xi \rightarrow \infty$, is plotted for comparison as a solid line.
}
\label{fig:SOT_Criticalcurrent}
\end{figure}

\subsubsection{Limiting case: Absence of non-adiabatic torques}
The limit of neglecting the non-conservative contribution to SOTs corresponds to setting $\tau_{\mathrm{DL}} \rightarrow 0$ while keeping $\xi \tau_{\mathrm{DL}}$ finite, see Eq.~\eqref{eq:SOT_torque}.
This corresponds in Eqs.~\eqref{eq:halfdomainmomentaSOT} 
to $\tilde{\tau} \rightarrow 0$ and $\xi \tilde{\tau}$ being finite, such that the right hand sides of Eqs.~\eqref{eq:halfdomainmomentaSOT} vanish and the left hand sides are conserved. 
Comparing the conserved quantities at $x=0$ and $x\rightarrow \infty$ we obtain the following critical current,
\begin{equation}\label{eq:result_SOT}
j_c^{\textrm{SOT}} = \frac{2 e l}{\xi \hbar \theta_\mathrm{Hall}}\left(\lambda - \frac{D^2}{2 J}\right).
\end{equation}

In the absence of chiral interactions, the critical current depends predominantly on the anisotropy strength, while in a chiral system, it is determined by a combination of the DMI strength, the exchange interaction and the anisotropy strength.
Again, we notice that for $D\geq D_c= \sqrt{2 J\lambda}$, there is no static solution with an applied current as the system decays to the helicoidal state. 
We have confirmed our analytical solution by micromagnetic simulations, where the results are shown in Fig.~\ref{fig:SOT_Criticalcurrent}.

\subsubsection{Including non-adiabatic torques}
For the full model, considering both contributions to SOTs, Eq.~\eqref{eq:SOT_torque}, we proceed analogously to the STTs case. Integrating over the semi-infinite nanowire and solving for the current density, we obtain that $j_{c}^{\textrm{SOT}}$ increases for $\xi \rightarrow 0$. Moreover, the critical current is finite for $\xi = 0$, see Fig.~\ref{fig:SOT_Criticalcurrent}d). 
In the absence of DMI, the critical current is given by
\begin{equation}\label{eq:jcSOTbeta}
j_{c}^{\textrm{SOT}} = \frac{2e l J \lambda^2}{\hbar\theta_{\mathrm{Hall}}}\left(\frac{
E_{\textrm{hel}} -\xi + \sqrt{(E_{\textrm{hel}} -\xi)^2 + \frac{2}{J \lambda}E_{\mathrm{ani}}^2}}{E_{\mathrm{ani}}^2}\right),
\end{equation}
where $E_{\mathrm{ani}} = \lambda \int_{0}^{\infty}dx(1-m_{x}^2)$ is the anisotropy energy, for details see App.~\ref{app:jcrit}. Also, for the calculation of the profiles of the magnetic texture below $j_{c}^{\textrm{SOT}}$ see App.~\ref{app:dwprofile}.

Overall, we find that the critical currents for SOTs standard parameters are lower than those for STTs, and therefore SOTs should facilitate a domain wall creation.

%%%%%%%%%%%%%%%%%%%%%%%%%%%%%%%%%%%%%%%%%%%%%%%%%%%%%%%%%%%%%%%%%%%%%%%%%%%%%%%%%%
%%%%%%%%%%%% Section IV 
%%%%%%%%%%%%%%%%%%%%%%%%%%%%%%%%%%%%%%%%%%%%%%%%%%%%%%%%%%%%%%%%%%%%%%%%%%%%%%%%%%
\section{Long-range magnetic domain wall interaction}
\label{sec:DWInteraction}

\begin{figure}
\includegraphics[width=0.5 \textwidth]{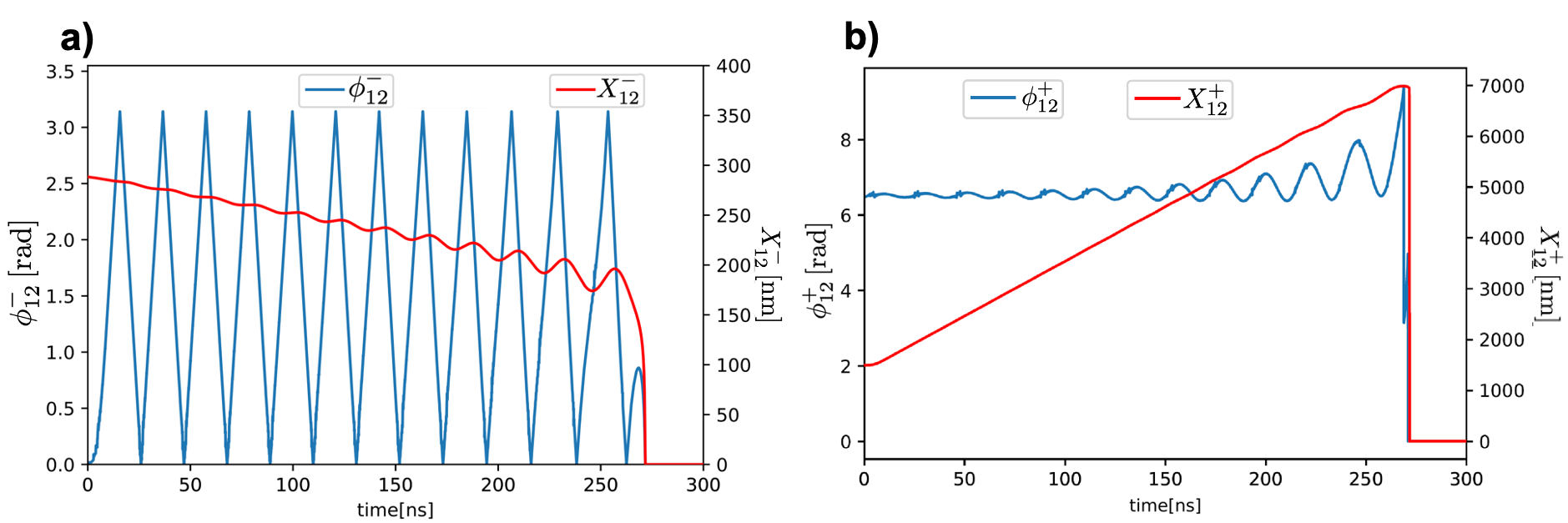}
\caption{Simulation results for the interaction of two domain walls moving in a wire showing their a) relative distance $X_{12}^{-}$ (relative angle $\phi_{12}^{-}$) on the right (left) $y$-axis, b) center of mass $X_{12}^{+}$ (total angle $\phi_{12}^{+}$) on the right (left) $y$-axis. Here we shifted $\phi_{12}^{+}$ by $2\pi$.
In this set-up, a tail-to-tail domain wall is chasing the slower head-to-head domain wall, so the distance shrinks in time. Notice that $\phi_{12}^{-}$ and $X_{12}^{+}$ has a rather constant slope, as shown in Eqs.~\eqref{eq:2DWsEOM}.
The yo-yo-like behavior of the two domain walls, reveals that the interaction is attractive or repulsive depending on the relative angle and distance between them, see Eq.~\eqref{eq:interaction_of_two_DW}. For this simulation we used a DMI strength of $D = 2.5\times10^{-5}
J/m^2$, damping parameters $\alpha = 0.5$, $\beta = 0$, and the current is $j = 2.5\times10^{11} A/m^2$. }
\label{fig:interaction}
\end{figure}

In chiral magnets, domain walls shedded by STTs annihilate in pairs upon traveling along the nanowire, see Fig.~\ref{fig:interaction}. This can be explained by two facts, both originated in the opposite chirality of consecutive injected domain walls: 1) consecutive domain walls have different velocities and therefore one is chasing another;~\cite{Tretiakov2010} 2) besides the translational motion of each domain wall, the magnetization of each of the walls also precesses. Consecutive domain walls, rotate in opposite directions such that at certain period of times it is easier for them to annihilate.

\begin{figure}
\includegraphics[width=0.5 \textwidth]{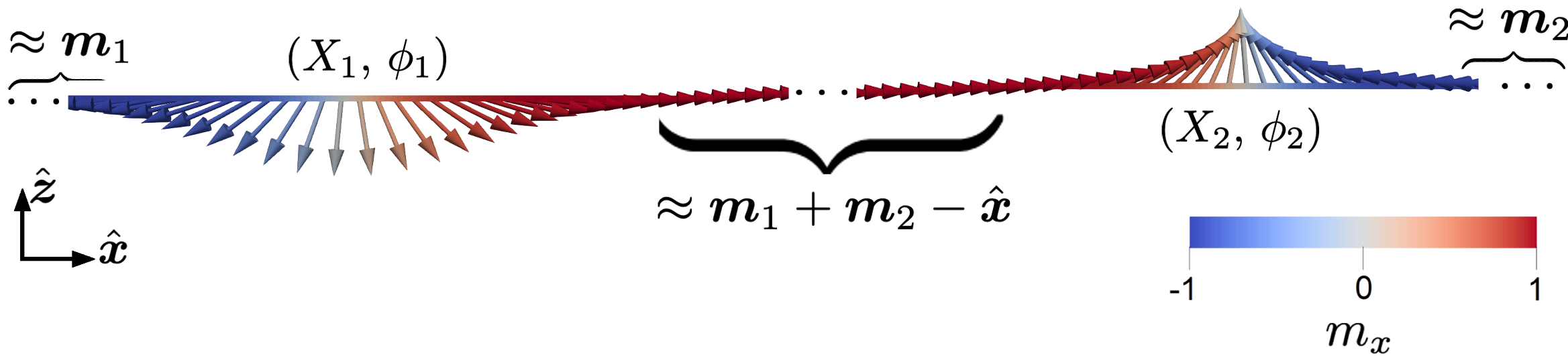}
\caption{Sketch of a pair of domain walls. To the left we have a tail-to-tail domain wall, described by $\vect{m}_{1}$, and to the right a head-to-head domain wall, described by $\vect{m}_{2}$. The magnetization between the domain walls can be approximated by the effective superposition of the two domain walls. At distances much bigger than the domain wall width, each domain wall can be considered as a rigid object described by the collective coordinates $X_{i}$ and $\phi_{i}$.}
\label{fig:dwpair}
\end{figure}

For our analysis, we simplify the description of the two domain walls to the case when they are far apart from each other, see Fig.~\ref{fig:dwpair}. In this case, i.e.\ when the 
distance between two domain walls is much bigger than their domain wall width, we consider the chiral domain walls as rigid objects. 
Each domain wall can be approximated by the solution of a single domain wall such that the following ansatz is a good approximation for the whole magnetic configuration 

\begin{equation}
\vect{m} \approx 
\begin{cases}
\vect{m}_{1} \quad & \text{if} \ x\lesssim X_{1}\\
\vect{m}_{1} + \vect{m}_2 - \hat{\vect{x}} \quad & \text{if} \ X_{1}\lesssim x\lesssim X_{2}\\
\vect{m}_{2} \quad & \text{if}\  x\gtrsim X_{2},
\end{cases}
\label{eq:parametrozation_of_m}
\end{equation}
with~\cite{Tretiakov2010}
\begin{align}\label{eq:parametrization_DW_interaction}
\vect{m}_{i} =& \pm \tanh[(x - X_{i})/\Delta]\hat{\vect{x}} \\
&+\frac{\cos\left[\Gamma (x - X_{i}) + \phi_{i}\right]\hat{\vect{y}} + \sin\left[\Gamma (x - X_ {i}) + \phi_{i}\right]\hat{\vect{z}}}{\cosh\left[(x - X_{i})/\Delta\right]} \notag
\end{align}
and $\Gamma = D/J$. The positive sign labels the domain wall with $i=1$ being a tail-to-tail domain wall, and the minus sign corresponds to $i=2$ being a head-to-head domain wall. Furthermore, $X_i$ are the center positions of the domain walls, with their relative distance $X_{12}^{-} \equiv X_{2} - X_{1} \gg \Delta$, and $\phi_i$ are their azimuthal angles.

Using this ansatz, we obtain from Eq.~\eqref{eq:energy}, the interaction energy between the domain walls
\begin{equation}
E_{12} \approx - \frac{J X_{12}^{-}\, e^{\frac{-X_{12}^{-}}{\Delta}}}{\Delta^2}\left(\Gamma^2\Delta^2 + 1\right) \cos\left(\Gamma X_{12}^{-} - \phi_{12}^{-}\right),
\label{eq:interaction_of_two_DW}
\end{equation}
where $\phi_{12}^{-}= \phi_2 - \phi_1$ the phase difference of the two domain walls, which we can define in between $0$ and $\pi$. 
Notice that the domain walls rotate in opposite directions.
The magnitude of the interaction energy decays exponentially as a function of the distance between the two domain walls. 
Moreover the interaction is attractive or repulsive depending on the interplay of $\phi_{12}^{-}$ and $X_{12}^{-}$.

We describe the dynamics of the domain wall pair by considering them as rather rigid objects. Within the collective coordinate approach,~\cite{Schryer1974,Tretiakov2008,Rodrigues2017} we obtain the following equations of motion (for the derivation, see App.~\ref{sec:CollectiveCoordinate}.)
\begin{subequations}\label{eq:dwsmotion}
\begin{align}
\dot{X}_{i} &= \pm  \frac{\gamma}{M_{s}}\frac{\partial \left(E_{12}+E_{\textrm{STT}}\right)}{\partial \phi_{i}} + \gamma_{X_{i}},\\
\dot{\phi}_{i} &= \mp  \frac{\gamma \Delta}{M_{s}}\frac{\partial  \left(E_{12}+ E_{\textrm{STT}}\right) }{\partial X_{i}}  + \gamma_{\phi_{i}},
\end{align}
\end{subequations}
where $E_{12}$ is the interaction energy between the domain walls and $E_{\mathrm{STT}} = - v\phi_{12}^{-}$.
From these we obtain the following equations of motion for the total and relative collective parameters,
\begin{subequations}
\begin{align}\label{eq:2DWsEOM}
\dot{X}_{12}^{+} =& \frac{2}{\Delta^2}\frac{\left(v \Delta^2 - \tilde{J} e^{-X_{12}/\Delta}\sin\left(\Gamma X_{12}^{-} - \phi_{12}^{-}\right)\right)}{(1 + \alpha^2)},\\
\dot{X}_{12}^{-} =& \frac{2\alpha}{\Delta^2}\frac{\left(-v \Gamma\Delta^3 + \tilde{J} e^{-X_{12}/\Delta}g \sin\left(\Gamma X_{12}^{-} - \phi_{12}^{-}\right)\right)}{(1 + \alpha^2)}.
\end{align}
\begin{align}
\dot{\phi}_{12}^{+} =& \frac{2 \tilde{J} g \, e^{\frac{-X_{12}^{-}}{\Delta}}}{\Delta^3}\left(\cos\left(\Gamma X_{12}^{-} - \phi_{12}^{-}\right)
+\Gamma\Delta \sin\left(\Gamma X_{12}^{-} - \phi_{12}^{-}\right) \right),\\
\dot{\phi}_{12}^{-} =& \frac{2\alpha g}{\Delta^3(1 + \alpha^2)}\Big(v\Delta^2 - \tilde{J} e^{-X_{12}^{-}}\Big(\Gamma\Delta\cos\left(\Gamma X_{12}^{-} - \phi_{12}^{-}\right)\notag\\
& - \sin \left(\Gamma X_{12}^{-} - \phi_{12}^{-}\right)\Big)\Big),
\end{align}
\end{subequations}
where $X_{12}^{+} = X_{1} + X_{2}$ corresponds to the center of mass, $\phi_{12}^{+} = \phi_{1} + \phi_{2}$ to the total phase, we define $\tilde{J} = J\gamma/M_{s}$, and $g = \left(\Gamma^2\Delta^2 + 1\right)$ is a constant that depends on the shape of the domain walls and we chose $\beta = 0$ for simplicity. We observe that the center of mass of the domain walls, $X_{12}^{+}$, moves forward at a rather constant speed with a small oscillation while the distance between the domain walls, $X_{12}^{-}$, decays with time with a small oscillation due to the interaction. The phase difference 
$\phi_{12}^{-}$ presents a similar behavior with a constant change plus an oscillation whose amplitude decays with the distance. We notice that $\Gamma\Delta = D/\sqrt{2J\lambda - D^2} \ll 1$ for small $D \ll D_{c}$.The total phase oscillates around $0$ due to the interaction. Overall, we observe an yo-yo-like approaching of the domain walls caused by the interaction of the domain walls and their relative constant velocity. Micromagnetic simulations corroborate with the analytical calculations, see Fig.~\ref{fig:interaction}.

%%%%%%%%%%%%%%%%%%%%%%%%%%%%%%%%%%%%%%%%%%%%%%%%%%%%%%%%%%%%%%%%%%%%%%%%%%%%%%%%%%
%%%%%%%%%%%% Section V
%%%%%%%%%%%%%%%%%%%%%%%%%%%%%%%%%%%%%%%%%%%%%%%%%%%%%%%%%%%%%%%%%%%%%%%%%%%%%%%%%%
\section{Discussion and Conclusion}
\label{sec:Conclusion}

In this paper we obtained the minimal electrical current to inject domain walls in semi-infinite chiral magnetic nanowires with a pinned magnetization at the end. We have demonstrated that it is possible to significantly reduce the necessary current by increasing the DMI. We also considered SOTs instead of STTs, which usually is associated to a more efficient manipulation of magnetic textures.~\cite{Miron2011,Ryu2013,Dao2019} Moreover, for the STTs case, we showed that the presence of non-adiabatic torques may reduce even further the critical current. While for the STTs one is able to periodically inject domain walls into the system, for SOTs, we showed that it is possible to only inject one domain wall with a fixed configuration.

In the periodic injection of domain walls by STTs, we notice that due to the presence of DMI, consecutive domain walls present different velocities and annihilate at some distance from the point of injection. This annihilation is preceded by a yoyo-like approximation of the domain walls due to their mutual interaction. The distance between them oscillates as the interaction alternates between an attractive and a repulsive potential depending on their relative distances and azimuthal angles. The annihilation of domain walls has been studied previously for non-chiral domain walls in Ref.~\onlinecite{Ghosh2017b}, where there is no oscillatory behavior.

We compared our analytical calculations to micromagnetic simulations and provided a solid understanding of the injection and interaction of chiral domain walls. The demonstrated decrease in the required electrical current density is essential for the design of more efficient domain wall based devices.

%%%%%%%%%%%%%%%%%%%%%%%%%%%%%%%%%%%%%%%%%%%%%%%%%%%%%%%%%%%%%%%%%%%%%%%%%%%%%%%%%%
%%%%%%%%%%%% Acknowledgments
%%%%%%%%%%%%%%%%%%%%%%%%%%%%%%%%%%%%%%%%%%%%%%%%%%%%%%%%%%%%%%%%%%%%%%%%%%%%%%%%%%
\section{Acknowledgments}

We acknowledge funding from the German Research Foundation (DFG), projects EV 196/2-1, TRR 173 - 268565370 (project B11),
EV196/5-1 and SI1720/4-1 as well as the Emergent AI Center funded by the Carl-Zeiss-Stiftung. 

This work was done while N.S. was working at Institute of Physics, Johannes Gutenberg University of Mainz, 55128 Mainz, Germany

%%%%%%%%%%%%%%%%%%%%%%%%%%%%%%%%%%%%%%%%%%%%%%%%%%%%%%%%%%%%%%%%%%%%%%%%%%%%%%%%%%
%%%%%%%%%%%% Appendix
%%%%%%%%%%%%%%%%%%%%%%%%%%%%%%%%%%%%%%%%%%%%%%%%%%%%%%%%%%%%%%%%%%%%%%%%%%%%%%%%%%
\appendix

\section{Inequality condition for the existence of stable static solutions for the half domain wall configuration}
\label{app:jcrit}
In this section, we present the inequalities that
allow for the calculation of the critical current in the general case.

%%%%%%%%%%%%%%%%%%%%%%%%%%%%%%%%%%%%%%%%%%%%%%%%%%%%%%%%%%%%%%%%%%%%%%%%%%%%%%%%%%
%%%%%%%%%%%% App. A
%%%%%%%%%%%%%%%%%%%%%%%%%%%%%%%%%%%%%%%%%%%%%%%%%%%%%%%%%%%%%%%%%%%%%%%%%%%%%%%%%%
\subsection{STTs}
\label{app:STTdetails}
For $\beta \neq0$ we integrate Eqs.~\eqref{eq:halfdomainmomentaSTT}  from $x=0$ to $x\rightarrow \infty$ 
and obtain
\begin{subequations}
\begin{align}
-J\partial m_{y}|_{x=0} - D \mp \frac{v M_{s}}{\gamma}&= \frac{\beta v M_{s}}{\gamma}E_{\textrm{hel}},\\
\frac{J}{2} \left((\partial m_{x})^2 + (\partial m_{y})^2\right)|_{x=0} - \lambda  &= \frac{\beta v M_{s}}{\gamma} \frac{2}{J} E_{\textrm{exch}},
\end{align}
\end{subequations}
where $E_{\textrm{exch}} = (J/2)\int_{0}^{\infty}dx\,(\partial\vect{m})^2$ is the exchange energy, and $E_{\textrm{hel}}= \int_{0}^{\infty} dx \, \hat{\vect{x}}\cdot(\vect{m}\times\partial\vect{m})$ the helicity energy.
Moreover, we used the boundary conditions  $\vect{m}(x=0) = \hat{\vect{z}}$, and at $x\rightarrow\infty$ we apply $\vect{m}= \hat{\vect{x}}$ and $\partial \vect m =0$. 
In analogy to the simple case, we now exploit, the general relation $\vect{m} \perp \partial\vect{m}$ to isolate $\partial m_{y}|_{x=0}$ in the first equation. Substituting this expression into the second equation, as well as, noticing that $(\partial m_{x})^2>0$, we obtain the following inequality
\begin{align}
\lambda - \frac{1}{2J}\Bigl(D + \frac{v M_{s}}{\gamma}\bigl(\pm 1 + \beta
E_{\textrm{hel}}\bigr)
\Bigr)^2 + \frac{\beta v M_{s}}{\gamma} 
\frac{2}{J} E_{\textrm{exch}}>0.
\end{align}
Notice that the total sign of $E_{\textrm{hel}}$ depends of the configuration at infinity, such that $E_{\textrm{hel}} = \pm |E_{\textrm{hel}}|$ for $\vect{m} = \pm \hat{\vect{x}}$ at infinity.
From which we obtain the critical value for $v$. For $D= 0$, the inequality above simplifies to
\begin{align}
\label{eq:A3}
\lambda - \frac{1}{2J}\Bigl(\frac{v M_{s}}{\gamma}\bigl(1 + \beta
|E_{\textrm{hel}}|\bigr)\Bigr)^2 + \frac{\beta v M_{s}}{\gamma} 
\frac{2}{J} E_{\textrm{exch}}>0.
\end{align}
At the critical current Eq.~\eqref{eq:A3} turns into an equality, which solving for the critical current leads to
Eq.~\eqref{eq:jcSSTbeta} from the main text.

\subsection{SOTs}
\label{app:SOTdetails}
In this section we present the full inequality that allows for the calculation of the critical current for SOTs in the general case. We follow the same steps as in the subsection above.
First we integrate Eqs.~\eqref{eq:halfdomainmomentaSOT} from $0$ to $\infty$ and obtain
\begin{subequations}
\begin{align}
&J\partial m_{y}|_{x=0} + D = - j^{\textrm{SOT}} \tilde{\tau} \frac{E_{ani}}{\lambda},\\
&\frac{J}{2}\left((\partial m_{y})^2 + (\partial m_{z})^2\right)|_{x=0}  - \lambda 
+ \xi j^{\textrm{SOT}} \tilde{\tau} =
 j^{\textrm{SOT}} \tilde{\tau} E_{\textrm{hel}},
\end{align}
\end{subequations}
where we introduced $\tilde{\tau}= \tau_{\mathrm{DL}} M_{s}/\gamma$ to shorten the notation and we
used that $\vect{m} \perp \partial\vect{m}$, $\vect{m}(x=0) = \hat{\vect{z}}$, and $\vect{m}(x\rightarrow\infty) = -\hat{\vect{x}}$. 
Analogously to the case above, we obtain the following inequality,
\begin{align}
\lambda + (E_{\textrm{hel}} -  \xi ) j^{\textrm{SOT}} \tilde{\tau} - \frac{1}{2J}
\left(D + j^{\textrm{SOT}} \tilde{\tau}
\frac{E_{ani}}{\lambda}
\right)^2 
>0.
\end{align}
From which we obtain the critical value for $j_{c}$. If we consider $D=0$ we obtain the simplified inequality,
\begin{align}
\lambda 
+ j^{\textrm{SOT}} \tilde{\tau} \left(
E_{\textrm{hel}} - \xi \right)
- \frac{1}{2J}\left(j^{\textrm{SOT}} \tilde{\tau} \frac{E_{ani}}{\lambda} \right)^2 >0,
\end{align}
from which the roots yield the critical current, i.e.\ Eq.~\eqref{eq:jcSOTbeta} from the main text.

\section{Magnetic profile below the critical current}
\label{app:dwprofile}
In this appendix we calculate explicitly the magnetization profile in the presence of STTs and SOTs below the critical current in the semi-infinite nanowire setup of the main text. For this, we define the gradient of the magnetization as
\begin{equation}\label{eq:GammaLambda}
\partial\vect{m} \equiv \Gamma(x)\hat{\vect{x}}\times\vect{m} + \Lambda(x)\vect{m}\times(\hat{\vect{x}}\times\vect{m}),
\end{equation}
where $\Gamma(x)$ and $\Lambda(x)$ define the shape of the magnetic configuration.
 We have therefore that
\begin{subequations}
\begin{align}
(\partial \vect{m})^2 &= \left(\Gamma^2 + \Lambda^2\right)(1-m_{x}^2),\\
\hat{\vect{x}}\cdot(\vect{m}\times\partial\vect{m}) &= \Gamma(1-m_{x}^2)
\end{align}
\end{subequations}
Notice that $|m_{x}|\leq1$.

\subsection{STTs}
To calculate the magnetic profile for STTs with $\beta =0$ we substitute \eqref{eq:GammaLambda} into Eqs.~\eqref{eq:halfdomainmomentaSTT} and obtain 
\begin{subequations}
\begin{align}
(J\Gamma - D)(1 + m_{x}) + \frac{v M_{s}}{\gamma} &=0,\\
\Gamma^2 + \Lambda^2 &= \frac{2\lambda}{J}.
\end{align}
\end{subequations}
Substituting $\Gamma$ from the first equation into the second one and using $\Lambda = \partial m_{x}/(1-m_{x}^2)$, we get
\begin{equation}
\frac{(\partial m_{x})^2}{(1-m_{x}^2)^2} = \frac{2\lambda}{J} - \frac{1}{J^2}\left(\frac{v M_{s}}{\gamma(1+m_{x})} - D\right)^2.
\end{equation}
The static half-domain wall profile in the presence of STTs, with applied current densities below the critical one, can be obtained by integrating above equation over the nanowire starting at infinity up to the position $x'$ 
\begin{align}
\int_{0}^{m'_{x}}\frac{d m_{x}}{(1-m_{x}^2)\sqrt{\left(2J \lambda - \left(\frac{v M_{s}}{\gamma(1+m_{x})} - D\right)^2\right)}} = \frac{x'}{J^2}.
\end{align}
Solving this integral equation for $m_x$ yields the profile of the magnetic texture below the critical current.

\subsection{SOTs}
Analogously, for SOTs with vanishing non-adiabatic torques, we obtain from Eqs.~\eqref{eq:halfdomainmomentaSOT},
\begin{subequations}
\begin{align}
\Gamma &= \frac{D}{J},\\
\frac{J}{2}\bigl(\Lambda^2 + \frac{D^2}{J^2} - \lambda\bigr)(1 - m_{x}) + \xi \tilde{\tau} j &= 0.
\end{align}
\end{subequations}
From these expressions we then obtain the magnetic texture's profile in the presence of SOTs for spin-currents below the critical current density \begin{equation}
\int_{0}^{m'_{x}}\frac{d m_{x}}{(1-m_{x}^2)\sqrt{\left(\lambda - \frac{D^2}{J^2}\right)(1-m_{x}) - \frac{2\xi \tilde{\tau} j}{J}}} = x'.
\end{equation}

%%%%%%%%%%%%%%%%%%%%%%%%%%%%%%%%%%%%%%%%%%%%%%%%%%%%%%%%%%%%%%%%%%%%%%%%%%%%%%%%%%
%%%%%%%%%%%% App. B
%%%%%%%%%%%%%%%%%%%%%%%%%%%%%%%%%%%%%%%%%%%%%%%%%%%%%%%%%%%%%%%%%%%%%%%%%%%%%%%%%%

\section{Collective coordinate approach}
\label{sec:CollectiveCoordinate}

In this section we provide the details for the dynamics of two interacting domain walls described by, Eqs.~\eqref{eq:dwsmotion} in the main text.
Let us first consider a single domain wall which moves as a rigid object, i.e.\
\begin{equation}
\label{eq:mdot}
\dot{\vect{m}} = -\dot{X}\partial\vect{m} \pm \dot{\phi}\hat{\vect{x}}\times\vect{m}\\
\end{equation}
where, as in the main text, $X$ and $\phi$ are the position and azimuthal angle of the magnetization at the center of the domain wall. 
The $\pm$ sign characterizes the type of domain wall, tail-to-tail (+) or head-to-head (-). In the following we consider the case, where $\dot{X}$ is positive for both domain walls. The tail-to-tail domain wall, however, is faster then the head-to-head domain wall, and their center magnetizations rotate in opposite directions.

The dynamical parameters $X$ and $\phi$ are called collective coordinates and are conjugated by a Poisson bracket for which we can describe a Hamiltonian formalism.~\cite{Rodrigues2017} Within this formalism we can easily take into account the interaction between two domain walls. In order to obtain the Poisson brackets of the collective coordinates, we need to derive the Berry phase dependence on the collective coordinates $X,\phi$ for the domain wall profile given in Eq.~\eqref{eq:parametrization_DW_interaction}. In the spherical coordinates, $\vect{m} = \sin\theta\left(\cos\varphi\hat{\vect{y}} + \sin\varphi\hat{\vect{z}}\right) + \cos\theta\hat{\vect{x}}$, the Berry phase is given by
\begin{equation}
\mathcal{S}_{B} = \frac{M_{s}}{\gamma}\int dx (1 - \cos\theta(X,\phi) )\dot{\varphi}
\end{equation}
Substituting $\cos\theta = \pm \tanh\left(x - X/\Delta\right)$ as well as $\varphi = \Gamma\left(x - X\right) + \phi$, and expanding on $X$ and $\phi$ we obtain
\begin{align}
\mathcal{S}_{B} \approx \mp \frac{\gamma}{M_{s} \Delta}X\dot{\phi}
\end{align}
as the leading order term in terms of the collective coordinates. This implies that the equations of motion are given by
\begin{subequations}
\label{eq:C4}
\begin{align}
\dot{X} &= \pm\frac{\gamma}{M_{s}}\frac{\partial E (X,\phi)}{\partial \phi} + \gamma_{X},\\
\dot{\phi} &= \mp\frac{\gamma\Delta}{M_{s}}\frac{\partial E (X,\phi)}{\partial X} + \gamma_{\phi}.
\end{align}
\end{subequations}
where $E$ is the total energy of the system and
\begin{subequations}
\begin{align}
\gamma_{X} &= \mp \left(\beta v\Gamma\Delta + \frac{\gamma \alpha}{M_{s}} (\dot{\phi}\Delta - \dot{X}\Gamma\Delta)\right),\\
\gamma_{\phi} &= \pm \frac{\gamma}{M_{s}}\left(\frac{\Gamma^2\Delta^2 +1}{\Delta}\left(\alpha\dot{X} - \frac{\beta v M_{s}}{\gamma}\right)) - \alpha\dot{\phi}\Gamma\Delta\right),
\end{align}
\end{subequations}
are the dissipation terms.

The total energy for the current-driven pair of rigid domain walls is given by the interaction between the domain walls and a term due to STTs,
\begin{equation}\label{eq:total_energy}
E(X_{1},X_{2},\phi_{1},\phi_{2}) = E_{12}(X_{1},X_{2},\phi_{1},\phi_{2}) + E_{\mathrm{STT}},
\end{equation}
with $E_{\mathrm{STT}} = M_{s}v(\phi_{1} - \phi_{2})/\gamma$.
Substituting the energy \eqref{eq:total_energy} into the equations of motion, Eqs.~\eqref{eq:C4}, and solving the system for $X_{i},\phi_{i}$ while setting $\beta = 0$ leads to Eqs.~\eqref{eq:dwsmotion} of the main text.

\end{document}